\documentclass[aps,prl,twocolumn,showpacs,showkeys,footinbib,amsmath,amssymb,superscriptaddress]{revtex4-2}

\usepackage{amsmath}
\usepackage{amssymb}
\usepackage{graphicx}
\usepackage{bm}
\usepackage{color}
\usepackage{relsize}
\usepackage{braket}
\usepackage{bbold}
\usepackage{txfonts}
\usepackage{epstopdf}
\PassOptionsToPackage{colorlinks=true,
	linkcolor=blue,
	citecolor=blue,
	urlcolor=blue}{hyperref}
\usepackage{hyperref}

\begin{document}
	
\title{Altermagnetic Proximity Effect }

\author{Ziye Zhu}
\affiliation{Eastern Institute for Advanced Study, Eastern Institute of Technology, Ningbo, Zhejiang 315200, China}

\author{Richang Huang}
\affiliation{Eastern Institute for Advanced Study, Eastern Institute of Technology, Ningbo, Zhejiang 315200, China}

\author{Xianzhang Chen}
\affiliation{Eastern Institute for Advanced Study, Eastern Institute of Technology, Ningbo, Zhejiang 315200, China}

\author{Zhou Cui}
\affiliation{Eastern Institute for Advanced Study, Eastern Institute of Technology, Ningbo, Zhejiang 315200, China}

\author{Xunkai Duan}
\affiliation{Eastern Institute for Advanced Study, Eastern Institute of Technology, Ningbo, Zhejiang 315200, China}

\author{Jiayong Zhang}
\affiliation{Eastern Institute for Advanced Study, Eastern Institute of Technology, Ningbo, Zhejiang 315200, China}
\affiliation{School of Physical Science and Technology, Suzhou University of Science and Technology, Suzhou, 215009, China}

\author{Igor \v{Z}uti\'c}
\affiliation{Department of Physics, University at Buffalo, State University of New York, Buffalo, New York 14260, USA}

\author{Tong Zhou}
\email{tzhou@eitech.edu.cn}
\affiliation{Eastern Institute for Advanced Study, Eastern Institute of Technology, Ningbo, Zhejiang 315200, China}	
\date{\today}

\vspace{1em}

\begin{abstract}
	Proximity effects complement conventional materials design by enabling interfacial properties absent in any constituent. Here we uncover an altermagnetic proximity effect (AMPE), distinct from ferromagnetic and antiferromagnetic proximity, in which the hallmark momentum-alternating spin splitting of an altermagnet is transferred across an interface into an adjacent nonmagnetic layer—a process we term altermagnetization. Using first-principles calculations and model analysis, we identify AMPE in heterostructures based on the prototypical van der Waals altermagnet V$_2$Se$_2$O, where a proximitized monolayer PbO acquires altermagnetic band splitting and real-space spin textures, with systematic tunability via interlayer spacing and magnetic configuration. We further demonstrate that AMPE enables valley-dependent spin splitting in the semiconductor PbS and realizes  topological superconductivity in the s-wave superconductor NbSe$_2$, both inheriting the altermagnetic spin texture. Finally, we validate the generality and experimental feasibility of AMPE by realizing it in a broader class of established altermagnets, including V$_2$Se$_2$O derivatives, Ruddlesden-Popper perovskites, and the metallic CrSb. Our results identify AMPE as a universal proximity mechanism and a versatile platform for engineering emergent quantum phenomena in heterostructures.
	
\end{abstract}

\maketitle

Proximity effects allow a material to acquire properties of its neighbors, becoming magnetic, superconducting, topologically nontrivial, or exhibiting enhanced spin–orbit coupling~\cite{Zutic2019, RevModPhys.77.935, PhysRev.187.580}. Magnetic proximity effects, particularly in ferromagnets, are widely used to generate spin splitting and tune magnetic parameters including anisotropy, coercivity, and exchange bias~\cite{PhysRev.187.580, MANNA201461,Gmitra2015:PRB, PhysRevLett.114.016603, Zhao2017NN, PhysRevLett.119.127403, Zhong2020, Choi2023, Zhou2023NM}. Antiferromagnetic proximity, in turn, relies on interfacial hybridization without fringing fields, enabling ultrafast spin dynamics and symmetry-enforced transport~\cite{Zutic2019,RevModPhys.90.015005,Shao2024:npjspintronics}. These mechanisms underpin modern spintronics, valleytronics, topological states, and heterostructures of superconductors and magnets~\cite{zutic2004:RMP,Schaibley2016:NRM,PhysRevLett.112.116404, Cai2023:AQT, Fu2008PRL}.

A growing class of unconventional magnets~\cite{WuPRB2027,Hayami2019momentum, Yuan2020:PRB,Libor2020Crystal,mazin2021prediction,yuan2021:prediction, JunweiliuNC2021, zhu2024observation, Liu2025}, often termed altermagnets (AM)~\cite{Vsmejkal2022:beyond, VSmejkal2022:emerging, Bai2024, song2025altermagnets, Krempasky2024:MnTe, SongAMNature2025, Zhang2025,Jiang2025,Sheoran2025:PRB,XDH2025prl,Bhowal2025Review,Cao2025PRL}, exhibits properties beyond common ferromagnets and antiferromagnets, most notably momentum-dependent alternating spin splitting without net magnetization. This nonrelativistic spin splitting and its tunability expand the opportunities in spintronics~\cite{jungwirth2025altermagnetic, Shao2021, Takagi2025:NM, Chen2025Unconventional,Liu2026:PRL}, multiferroics~\cite{Duan2024PRL, zhu2025two, zhu2025emergent, Gu2024PRL, sun2025proposing, cao2024designing,guo2025altermagnetic,UrruPRB2025,PengNPJQM2025, Cui2026arXiv, Zhu2025AMSFET}, topotronics~\cite{PhysRevB.110.064426, Guo2023npj, feng2025type, YZM2025, PhysRevB.110.224436, Chen2026arXiv}, and superconductivity~\cite{brekke2023two, Li2023Majorana, zhu2023topological, ghorashi2024altermagnetic, Ouassou2023:PRL,Zhang2024:NC}. A central challenge now lies in understanding if this distinctive magnetic order, rooted in antiferromagnetic (AFM) sublattices linked by symmetries involving rotations~\cite{Vsmejkal2022:beyond, VSmejkal2022:emerging, Bai2024, song2025altermagnets} can be transferred across an interface into otherwise nonmagnetic (NM) layers. Establishing whether AM generate a distinct proximity effect, rather than simply mimicking ferromagnetic or antiferromagnetic proximity, is key to assessing their impact. Equally important is clarifying if this proximity channel can enable unexplored phenomena, particularly in van der Waals (vdW) heterostructures, where proximity effects are most tunable~\cite{Lazic2016:PRB,Xu2018:NC,FariaJunior2023:2DM,Zollner2023:PRB,Huang2024:PRB}.

\begin{figure}[t!]
	\centering
	\vspace{0.2cm}
	\includegraphics*[width=0.45\textwidth]{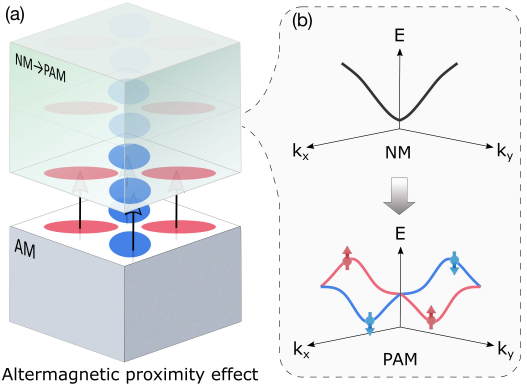}
	\vspace{-0.2cm}
	\caption{
		Schematic of the AMPE. {\bf (a)} Altermagnetism penetrating an NM layer. {\bf (b)} Band evolution of the NM layer as it becomes a proximitized altermagnet (PAM).
	}
	\label{Figure1}
\end{figure}

\begin{figure*}[t!]
	\centering
	\includegraphics[width=0.9\textwidth]{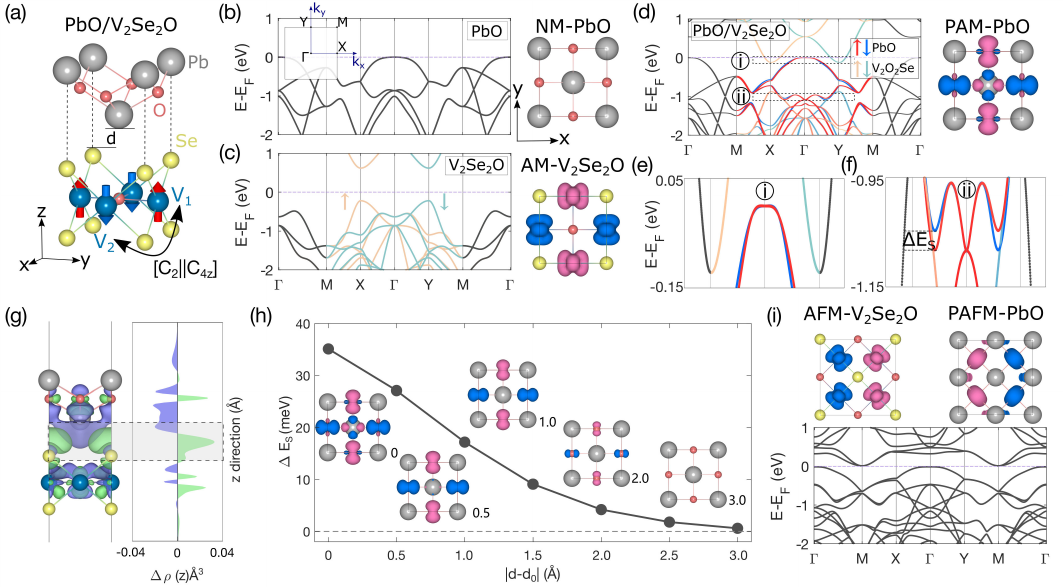}
	\caption{
		{\bf (a)} Crystal structure of PbO/V$_2$Se$_2$O heterostructure. The spin sublattices of V$_2$Se$_2$O are connected by $\left[C_2 || C_{4z}\right]$ symmetry. {\bf (b)}-{\bf (d)} Calculated bands and spin densities of pristine monolayers PbO and V$_2$Se$_2$O, and their heterostructure. The inset in {\bf (b)} shows the first Brillouin zone. {\bf (e)} and {\bf (f)} Enlarged views of the regions labeled \romannumeral 1~and \romannumeral 2~in {\bf (d)}. {\bf (g)} Differential charge density of the PbO/V$_2$Se$_2$O heterostructure and its planar average, $\Delta q(z)$. Green (purple) indicates charge accumulation (depletion) and the isosurface value is $5 \times 10^{-4}$~e/bohr$^3$. {\bf (h)} Spin splitting $\Delta E_S$ in {\bf (f)} with the spin densities, as a function of the interlayer distance $d$ in {\bf (a)}, where $d_0$ is the equilibrium value. {\bf (i)} Same as {\bf (d)} but for a proximitized AFM PbO (PAFM-PbO), realized when PbO is placed on a $\sqrt{2} \times \sqrt{2}$ V$_2$Se$_2$O supercell in a conventional AFM state. For all bands, black denotes spin-degenerate states; red and blue indicate spin-up and spin-down bands of the PAM component (PbO), while yellow and green represent the corresponding bands of the V$_2$Se$_2$O component.
	}
	\label{Figure2}
\end{figure*} 

In this work, we address these questions by introducing the altermagnetic proximity effect (AMPE) as in Fig.~\ref{Figure1} in which the momentum-alternating spin splitting intrinsic to AM is imprinted onto a neighboring nonmagnetic layer through an interfacial process we term altermagnetization. Combining first-principles calculations with model analysis, we establish AMPE in heterostructures formed from the prototypical AM V$_2$Se$_2$O and monolayer PbO: the PbO layer becomes altermagnetized, as evidenced by the induced altermagnetic band splitting and real-space spin densities, whose systematic evolution with interlayer spacing and magnetic configuration demonstrates both robustness and tunability of AMPE. We then illustrate broader consequences of this effect by showing that AMPE can generate valley-dependent spin splitting in PbS and enable a topological superconductivity in NbSe$_2$. Finally, by extending our analysis beyond the V$_2$Se$_2$O platform to additional AM classes, including V$_2$Se$_2$O derivatives, Ruddlesden–Popper perovskites, and the metallic CrSb, we establish AMPE as a distinct and universal proximity mechanism and a viable route for engineering multifunctional quantum states in heterostructures.

Our proposed AMPE is schematically illustrated in Fig.~\ref{Figure1}(a). When an NM material is placed in contact with an altermagnet, the extension of electronic wave functions across the interface can imprint the characteristic alternating spin splitting of the AM onto the NM layer. This interfacial process drives a transition from the NM state to a proximitized altermagnet (PAM), in which originally spin-degenerate bands acquire momentum-dependent spin polarization [Fig.~\ref{Figure1}(b)]. As other proximity effects~\cite{Zutic2019}, AMPE is short ranged and decays with distance from the interface. Two-dimensional vdW systems, with atomically thin layers and clean interfaces, thus provide an ideal platform to exploit this AMPE and to design the corresponding highly tunable functionalities.

While most experimentally confirmed altermagnets exist as bulk crystals or thin films~\cite{Krempasky2024:MnTe, SongAMNature2025, Zhang2025,Jiang2025}, V$_2$Se$_2$O stands out as a famous vdW candidate~\cite{JunweiliuNC2021, PhysRevB.98.075132}. Its intercalated metallic derivatives have also been experimentally identified as altermagnets~\cite{Zhang2025,Jiang2025}, further highlighting its versatility. V$_2$Se$_2$O crystallizes in a square lattice where the two V sublattices form a checkerboard AFM configuration linked by $C_{4z}$ symmetry [Fig.~\ref{Figure2}(a)]. This symmetry yields the hallmark AM features, as confirmed by our calculations showing alternating spin densities in real space and momentum-dependent spin splitting [Fig.~\ref{Figure2}(c)]. To demonstrate the AMPE of V$_2$Se$_2$O, we select PbO as the adjacent NM layer. PbO is a well-established semiconductor with a simple, well-characterized crystal structure and spin-degenerate bands~\cite{barraza2021colloquium, kumar2018alpha}, as shown in our calculations [Fig.~\ref{Figure2}(b)]. Importantly, PbO and V$_2$Se$_2$O both form 2D square lattices with a minimal lattice mismatch of just 0.7\%, making PbO an ideal and clean reference to isolate and analyze the AMPE from V$_2$Se$_2$O.

To investigate the emergence of AMPE, we computed the total energies of several stacking configurations of the PbO/V$_2$Se$_2$O heterostructure (see Supplemental Material~\cite{SM}) and identified the most stable arrangement shown in Fig.~\ref{Figure2}(a). The corresponding band structure [Fig.~\ref{Figure2}(d)] indicates that both constituents largely preserve their intrinsic electronic features under vdW coupling. Strikingly, however, the PbO layer now exhibits pronounced momentum-dependent spin splitting [Figs.~\ref{Figure2}(d)-\ref{Figure2}(f)], in sharp contrast to its pristine spin-degenerate state [Fig.~\ref{Figure2}(b)], demonstrating that PbO is transformed into a PAM-PbO through AMPE. The induced spin splitting displays a characteristic AM signature: spin degeneracy is preserved along $\Gamma$–M, while opposite spin polarizations appear along M–X–$\Gamma$ and $\Gamma$–Y–M, consistent with the underlying order of V$_2$Se$_2$O. Moreover, the real-space spin density of PAM-PbO mirrors the symmetry of the V$_2$Se$_2$O substrate, providing direct evidence that PbO inherits AM character of V$_2$Se$_2$O. To confirm this, we further convert V$_2$Se$_2$O into a conventional AFM configuration by rearranging its spin sublattices [Fig.~\ref{Figure2}(i); see Supplemental Material~\cite{SM}]. In this case, the sublattices are related by translation or inversion symmetry, eliminating the alternating spin splitting. As expected, the induced splitting in PbO disappears [Fig.~\ref{Figure2}(i)], while its spin density develops an AFM feature, indicating a conventional AFM proximity effect. This control case further illustrates the AMPE of V$_2$Se$_2$O.

To further elucidate how PbO is influenced by the AMPE of V$_2$Se$_2$O, we analyze the interfacial interaction and charge redistribution in their vdW heterostructure. The planar-averaged charge density difference $\Delta q(z)$, together with the real-space redistribution shown in Fig.~\ref{Figure2}(g), indicates a slight interfacial charge transfer from PbO to V$_2$Se$_2$O. Such a modest charge transfer has a negligible effect on the AM ground state of V$_2$Se$_2$O, as further corroborated by the doping simulations in Supplemental Material~\cite{SM}. Because the strength of proximity effects is highly sensitive to interlayer spacing, we evaluate the evolution of the spin splitting $\Delta E_S$ as a function of the separation $d$ between PbO and V$_2$Se$_2$O. As shown in Fig.~\ref{Figure2}(h), the induced $\Delta E_S$ decreases monotonically with increasing $d$, accompanied by a corresponding reduction of the spin density, confirming the short-range nature of AMPE. Taken together, these results provide unambiguous evidence for AMPE from multiple, mutually consistent perspectives.

\begin{figure}[t!]
	\centering
	\includegraphics*[width=0.46\textwidth]{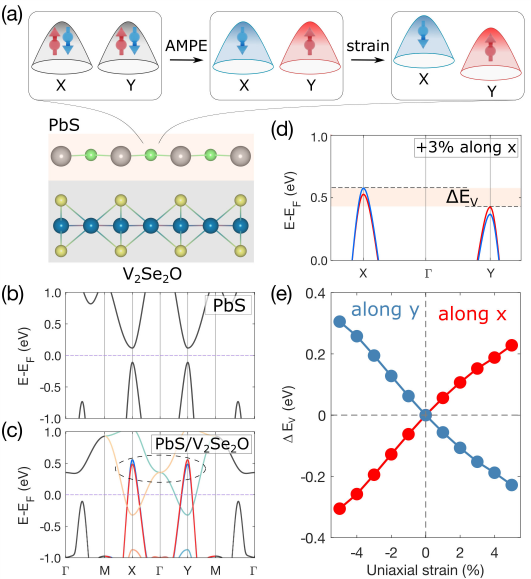}
	\caption{ 
		{\bf (a)} Schematic of the PbS/V$_2$Se$_2$O heterostructure illustrating the AMPE- and strain-induced spin and valley splitting in a monolayer PbS. {\bf (b)} and {\bf (c)} Calculated bands of the monolayer PbS and PbS/V$_2$Se$_2$O heterostructure. {\bf (d)} Bands of the AM-proximitized monolayer PbS under 3\% uniaxial tensile strain along the $x$ axis, showing valley splitting, $\Delta E_V$. {\bf (e)} Strain-dependent $\Delta E_V$. Band color codes as in Fig.~\ref{Figure2}.
	}
	\label{Figure3}
\end{figure}

\indent{$Valleytronics$}---
Beyond charge and spin, the valley degree of freedom provides a powerful route toward valleytronic devices and robust topological states~\cite{Xiao20227prl,Schaibley2016:NRM, Zhou2021:PRL, Zhou2015nano, Zhou2018npj}. A common strategy to lift valley degeneracy is to apply an external magnetic field, but the typically small $g$-factors of semiconductors limit its effectiveness. Magnetic proximity effects offer an alternative by introducing valley asymmetry~\cite{Zutic2019,Choi2023,Zhou2023NM}. Valley phenomena in altermagnets have recently attracted growing attention~\cite{JunweiliuNC2021,zhang2024predictable,Li2025Valley,PhysRevB.110.L220402}. In particular, monolayer V$_2$Se$_2$O has been predicted to host a unique symmetry-paired spin-valley locking that links spin and valley space with real space, enabling giant piezomagnetism and large noncollinear spin currents~\cite{JunweiliuNC2021}.

Motivated by this, we explore whether such valley physics can be transferred into a NM material through AMPE, using a PbS/V$_2$Se$_2$O heterostructure as an example [Fig.~\ref{Figure3}(a)]. Monolayer PbS, known for its rich topological, valleytronic, and optoelectronic properties~\cite{Wan2017}, exhibits spin-valley degeneracy in its pristine state [Fig.~\ref{Figure3}(b)]. When coupled to V$_2$Se$_2$O, PbS inherits the altermagnetic feature of symmetry-paired spin-valley locking through AMPE, as evidenced by the characteristic spin splitting in Fig.~\ref{Figure3}(c) and the corresponding rotation-symmetry-linked spin-density patterns shown in Supplemental Material~\cite{SM}. This pairing enables strain-tunable valley polarization: breaking the mirror symmetry between the X and Y valleys converts them from degenerate to polarized states. Uniaxial strain along $x$ or $y$ shifts their relative energies in opposite directions, producing a controllable valley splitting that grows monotonically under either compressive or tensile strain [Fig.~\ref{Figure3}(e)]. This effect is substantial even under moderate strain---for example, 3\% tensile strain along $x$ yields a 152~meV valence-band splitting [Fig.~\ref{Figure3}(d), Supplemental Material~\cite{SM}], far exceeding the thermal energy at 300$\,$K. Such a robust splitting ensures stable valley polarization and enables practical control of valley populations through mechanical or substrate engineering. These results demonstrate that AMPE provides a general route to induce spin-valley polarization in otherwise nonmagnetic materials. By combining interfacial coupling with strain engineering, one can realize electrically and mechanically tunable valley functionalities, opening new opportunities for spintronic and valleytronic applications in AM-based heterostructures.

\begin{figure}[t!]
	\centering
	\includegraphics[width=0.47\textwidth]{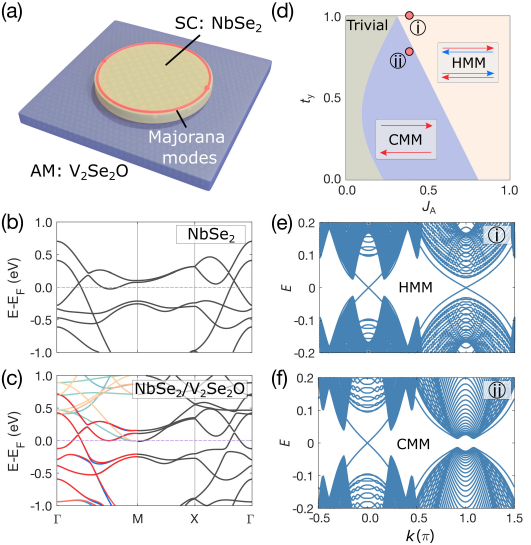}
	\caption{
		{\bf (a)} Schematic transformation of an $s$-wave superconductor NbSe$_2$ into a topological superconductor hosting edge MM (red circle) via AMPE from V$_2$Se$_2$O. {\bf (b)} and {\bf (c)} Bands of a NbSe$_2$ supercell and the NbSe$_2$/V$_2$Se$_2$O heterostructure. {\bf (d)} Topological phase diagram of the PAM-NbSe$_2$ obtained from Eq.~(\ref{eq1}). {\bf (e)} and {\bf (f)} Ribbon spectra of PAM-NbSe$_2$ showing helical and chiral MM (HMM and CMM). Parameters are $t_x=1$, $\mu=0.6$, $\lambda_R=0.2$, $J_A=0.4$, $\Delta=0.2$. $t_y=1$ in {\bf (e)} and $t_y=0.8$ in {\bf (f)}. Band color codes as in Fig.~\ref{Figure2}.
	}
	\label{Figure4}
\end{figure}

\indent{$Topological~superconductivity$}---
Proximity effects provide one of the most promising routes to realize topological superconductivity and host Majorana modes (MM), essential for fault-tolerant quantum computing~\cite{Flensberg2021:NRM,Sarma2015Majorana,Gungordu2022:JAP}. The conventional platform, an $s$-wave superconductor proximitized into a semiconductor with strong spin–orbit coupling under an external magnetic field, has yielded multiple signatures of topological superconductivity~\cite{Sarma2015Majorana}. However, this approach requires applied magnetic fields that compete with superconductivity. In contrast, AMPE offers a distinct advantage: it induces momentum-dependent spin splitting without net magnetization or external fields, thus preserving the superconducting gap while enabling new routes to engineer topological superconducting states~\cite{ghorashi2024altermagnetic,zhu2023topological,Li2023Majorana}.

To explore this possibility, we consider a platform composed of the well-studied vdW $s$-wave superconductor NbSe$_2$ placed on V$_2$Se$_2$O [Fig.~\ref{Figure4}(a)]. In this configuration, the AMPE-induced splitting from V$_2$Se$_2$O is expected to modify the pairing of NbSe$_2$ and can drive it into a topological superconductor hosting edge MM. Our first-principles calculations confirm that pristine NbSe$_2$ exhibits spin-degenerate bands [Fig.~\ref{Figure4}(b)], while in the NbSe$_2$/V$_2$Se$_2$O heterostructure these bands acquire altermagnetic spin splitting as seen in Fig.~\ref{Figure4}(c) and Supplemental Material~\cite{SM}, demonstrating that AMPE penetrates the superconducting layer without introducing net magnetization.

To analyze the superconducting properties of proximitized NbSe$_2$, we construct an effective model in the Nambu basis,
\begin{equation}
	\begin{aligned}
		H_{BdG} =&\left( t_x \cos{k_x}+t_y\cos{k_y} -\mu \right) \sigma _{0}\tau _{z}\\
		+&\lambda _{R}\left( \sin{k_{y}}\sigma _{x}\tau
		_{z}-\sin{k_{x}}\sigma _{y}\tau _{z}\right) \\
		+&J_A \left( \cos{k_x}-\cos{k_y} \right) \sigma _{z}\tau _{0}+\Delta \sigma
		_{0}\tau _{x},
	\end{aligned}
	\label{eq1}
\end{equation}
\noindent where $\sigma_{i}\left(\tau _{i}\right)$ are Pauli matrices in spin (particle–hole) space. The model includes kinetic hopping terms $t_{x/y}$, chemical potential $\mu$, $s$-wave pairing $\Delta$, the AMPE strength $J_A$, and the Rashba spin-orbit coupling strength $\lambda_R$, which arises from broken out-of-plane mirror symmetry at the interface and is essential for driving NbSe$_2$
into a topological superconducting phase~\cite{yi2022crossover,amundsen2024colloquium}, as detailed in the Supplemental Material~\cite{SM}. 

From gap-closing conditions and the calculated topological invariants~\cite{zhu2024field, SM}, we obtain the phase diagram in Fig.~\ref{Figure4}(d). For isotropic hopping ($t_x$ = $t_y$), crystalline symmetry enforces a valley degeneracy that, with suitable $J_A$, yields a helical topological superconductor hosting pairs of helical MM at the edges [Fig.~\ref{Figure4}(e)]. Breaking this crystalline relation ($t_x$ $\neq$ $t_y$) lifts the valley degeneracy, allowing valley-selective topological transitions: one valley becomes topological while the other remains trivial. The edge spectrum then contains a single chiral MM per boundary [Fig.~\ref{Figure4}(f)]. This symmetry-controlled switch between helical and chiral MM within a single platform is a key advantage of AMPE-based designs. 
Because altermagnetism can be flexibly tuned by various dynamic means~\cite{Bai2024,song2025altermagnets,Chen2025prl}, AMPE enables controllable manipulation of MM, offering unprecedented opportunities for their fusion and braiding~\cite{zhou2022fusion,Zhou2020prl,Fatin2016:PRL} central to topological quantum computing.

Beyond the V$_2$Se$_2$O-based heterostructures, we further demonstrate that AMPE is a generic interfacial mechanism by realizing it in heterostructures between other nonmagnetic layers and diverse AM platforms, including V$_2$Se$_2$O derivatives,  Ruddlesden–Popper perovskites (e.g., Ca$_3$Mn$_2$O$_7$~\cite{Gu2024PRL}), and experimentally established metallic altermagnet CrSb~\cite{SongAMNature2025}, as detailed in the Supplemental Material~\cite{SM}. These results further reinforce both the generality and experimental feasibility of AMPE.

Our proposed AMPE shifts the role of altermagnets from intrinsic or bulk mechanisms to an interfacial route. This conceptual advance significantly broadens their scope, enabling their momentum-alternating spin textures to be harnessed in otherwise NM systems. Just as ferromagnetic proximity can induce energy splittings beyond those achievable with external fields~\cite{Zutic2019}, AMPE offers comparable advantages while avoiding stray fields and net magnetization. This field-free distinction is especially crucial for proximity-induced topological superconductivity, where conventional Zeeman-based schemes are often limited to semiconductors with large-$g$ factors even under magnetic textures and fringing fields~\cite{Fatin2016:PRL,PhysRevB.99.134505}.

More broadly, the tunable spin and symmetry properties of altermagnets make AMPE a versatile platform for realizing emergent states that would otherwise demand complex materials or multiple proximity effects~\cite{Zutic2019}. A natural outlook is to identify the governing factors and control knobs of AMPE strength, as well as the characteristic energy and time scales for reconfiguring altermagnetic order. Such efforts will be essential for assessing the feasibility of dynamical control over spin-dependent and topological responses~\cite{zutic2004:RMP,amundsen2024colloquium}, opening new opportunities to manipulate quantum phases in both normal and superconducting states.

\begin{acknowledgements}
\indent{$Acknowledgements$}---We thank Yuntian Liu for helpful discussions. Tong Zhou acknowledges support from Hui Yang and inspiration from Avery Zhou and Mila Zhou. This work is supported by the Zhejiang Provincial Natural Science Foundation of China (LR25A040001), the Zhejiang Provincial Leading Innovative and Entrepreneurial Team Project (2025R01017), the National Natural Science Foundation of China (12474155, 12504108, and 11904250), the Zhejiang Pioneer Project under Grant No. 2025C06SA201986, the China Postdoctoral Science Foundation (2025M773440), the U.S. DOE, Office of Science BES, Award No. DE-SC0004890 (I.\v{Z}. for AM), and the U.S. ONR under Award No. MURI N000142212764 (I.\v{Z}. for MM). 

Z.Z. and R.H. contributed equally to this work.

\end{acknowledgements} 

\bibliography{AMPE_Refs_withSM}

\end{document}